# Binary and recycled pulsars: 30 years after observational discovery

G S Bisnovatyi-Kogan

Contents

1. Introduction
2. Magnetic fields and periods of radio pulsars
3. Binary X-ray pulsars
4. Theoretical prediction. Unavoidable formation of binary recycled radio pulsars
5. Magnetic field decay during accretion
6. Evolutionary formation schemes
7. Binary radio pulsars in globular clusters. The formation of single recycled pulsars
8. Relativistic effects in compact neutron star binaries
8.1 Observed relativistic effects
8.2 Relativistic effects in the Hulse ± Taylor pulsar
9. Two pulsars in a binary system: the most compact neutron star binary as the best laboratory for testing GR
9.1 Pulse arrival time in the binary pulsar
9.2 Results of data analysis [71]
9.3 Further prospects
9.4 The past and future of pulsars in a binary system
10. Binary neutron star coalescences: gravitational wave and gamma-ray bursts
10.1 Galactic binary neutron star coalescence rate and gravitational-wave pulse registration
10.2 Binary neutron star coalescences and gamma-ray bursts
11. Conclusion
References

**Abstract.** Binary radio pulsars, first discovered by Hulse and Taylor in 1974 [1], are a unique tool for experimentally testing general relativity (GR), whose validity has been confirmed with a precision unavailable in laboratory experiments. In particular, indirect evidence of the existence of gravitational waves has been obtained. Radio pulsars in binary systems (which have come to be known as recycled) have completed the accretion stage, during which neutron star spins reach millisecond periods and their magnetic fields decay 2 to 4 orders of magnitude more weakly than ordinary radio pulsars. Among about a hundred known recycled pulsars, many have turned out to be single neutron stars. The high concentration of single recycled pulsars in globular clusters suggests that close stellar encounters are highly instrumental in the loss of the companion. A system of one recycled pulsar and one 'normal' one discovered in 2004 is the most compact among binaries containing recycled pulsars [2]. Together with the presence of two pulsars in one system, this suggests new prospects for further essential improvements in testing GR. This paper considers theoretical predictions of binary pulsars, their evolutionary formation, and mechanisms by which their companions may be lost. The use of recycled pulsars in testing GR is discussed and their possible relation to the most intriguing objects in the universe — cosmic gamma-ray bursts — is examined.

## 1. Introduction

The discovery of pulsars in 1967, along with quasars and cosmic microwave background radiation discovered several years earlier, was the greatest milestone in physics and astronomy. Unlike the quasars and the cosmic microwave background, which had been dedicatedly studied before their discovery, pulsars were discovered quite accidentally while investigating interplanetary radio scintillations, which unexpectedly turned out to be strictly periodic. Their periods are kept constant to within more than the fifth digit after the decimal point. Observations of the scintillations started in July 1967, and at the end of August 1967, Jocelyn Bell, then a post-graduate student of Prof. Hewish from the Mullard Radio Observatory in Cambridge, UK, reported to her supervisor the discovery of an unusually rapidly variable radio source, apparently located far away from the solar system. The properties of the source were so intriguing that this discovery was not announced even at the greatest astronomical forum, the 13th IAU General Assembly, which took place in Prague in September 1967. Only after a thorough analysis, which firmly established the existence of a radio source with unprecedented period stability, did a paper in *Nature* announce the discovery of this radio source with the period $P = 1.3372795 \pm 0.0000020$ s [3], originally called CP 1919.

After my post-graduate studies at the Moscow Institute of Physics and Technology in 1967, I continued to work in the Institute of Applied Mathematics in the department headed by Ya B Zel'dovich. Zel'dovich trained his collaborators to be up on events, to continually study the scientific literature, which gave a lot of food for thought. In addition, he always forced us to think over what 'actually' occurs out there (in the cosmos). As a result of such a life 'under the press,' somewhere at the end of 1972, I suddenly realized that close binary radio pulsars must necessarily exist in nature, which at that time nobody could discover. We then lived in the epoch of great astronomical discoveries: quasars, cosmic microwave background, and binary X-ray sources discovered by the UHURU satellite. In the course of my post-graduate education, Zel'dovich charged me with studying stars, and therefore, among these new discoveries, I carefully investigated pulsars. I have a notebook where I made a summary of about 100 first papers on pulsars. In addition to problems related to the radiation mechanisms and huge brightness temperature of pulsars, their being solitary was intriguing, especially because a good half of their progenitor stars must have been binaries. Two explanations for that appeared most likely. The first was suggested by many authors and used the fact that during the formation of a pulsar in the supernova explosion, such a great amount of matter is expelled that the binary system should be disrupted, as was originally proposed by H Blaauw. The second explanation was suggested by V F Shvartsman in paper [4], eloquently entitled "Neutron stars in binary systems should not be pulsars." The idea was that radio emission must be absorbed by matter accreting into a neutron star. In fact, in that paper, the existence of X-ray pulsars in binary systems was predicted, which shortly after was discovered by UHURU. Shvartsman also predicted that "should pulsars be discovered in binaries, they must predominantly have wide orbits." This prediction was also partially confirmed by the discovery of a radio pulsar in a wide orbit around a B-star [5] (the orbital period is 1130 d, the pulsar period is 47.8 ms), although many more pulsars have been found to reside in close binaries.

The conclusion that pulsars in close binaries should necessarily exist followed directly from observations [6] and was made after the optical identification of the X-ray pulsar Her X-1. In that system, the companion to the X-ray pulsar is a relatively low-mass star, with the mass about two solar ones. Not only could such a star not lose a large amount of mass during the explosion, it could not explode at all and would ultimately evolve into a white dwarf after losing about half its mass via stellar wind outflow. After that, the mass transfer would stop and the neutron star would appear again as a radio pulsar. Estimates showed that at the stage of an X-ray Hercules-type pulsar, the rotation of the neutron star should accelerate such that the re-born pulsar (later called recycled) should be rotating very rapidly, with a period of less than one second. The evolutionary time of the optical star in Her X-1 is about 10 mln years, and hence many massive stars in binaries should have been turned into pulsars in close binaries over the course of the universe's existence. I thought that the only reason why this is not observed is the weakness of radiation from pulsars that have passed the accretion stage, and this could be related only to their anomalously small magnetic fields, which could have been screened by accreting plasma. I shared these ideas with B V Komberg, with whom I then collaborated on constructing a model for Her X-1. From discussions with him, it also became clear that binary systems containing X-ray pulsars with massive companions like the source in Centaurus, after completing evolution and the explosion of the secondary star as a supernova, could be destroyed to form two spatially close pulsars exhibiting certain correlations between ages, periods, and distance. As a result, we wrote a joint paper [6] in which we presented these considerations and estimates. We also included the result of analysis of pulsar data, carried out by Komberg, which included 10 pairs of close (single) pulsars that could have a joint origin from the same binary progenitor. Now such a joint origin appears unlikely because no second pulsar is observed in most binary pulsars with the secondary neutron star. The neutron star spin axes in binaries may probably become unparallel after collapses and explosions, as well as due to the loss of the rotational energy via pulsar radiation, and we can typically see only one pulsar beam. However, there is no restriction on both beams accidentally shining on the earth, and double radio pulsars are possible [7]. A close binary system of two pulsars was discovered only in 2004 [2]. Some of the pairs proposed by us in Ref. [6] may have a joint origin. Anyway, the main conclusion in our paper was that observations combined with the theory of stellar evolution "do not just allow excluding the possibility of the existence of a neutron star in pair with another star, which is also at the final evolutionary stage (white dwarf, neutron star, collapsar), but, moreover, make such a conclusion very probable. Therefore, the absence of radio pulsars in close binaries (as of April 27, 1973 — $G\,B$) requires ... an additional assumption about the magnetic field decay." In the fall of 1974, about half a year after our paper was published, Prof. D Pines visited Moscow and reported the sensational discovery of the first binary pulsar PSR 1913+16 by R Hulse and J Taylor. He spoke about this during a meeting with Zel'dovich and a part of our group at the Institute for Physical Problems and noted that the second (according to the period of rotation) pulsar found in this system had to be very young. I remember my immediately having stood up and told him about our paper. I argued that the newly discovered pulsar should be old and should have a weak magnetic field. Prof. Pines listened to me condescendingly and then repeated that according to the opinion of the leading US experts, the pulsar is young because it is spinning very rapidly. He did not react to my attempts to explain that the rapid rotation is the result of acceleration during accretion, and the talk changed to a different subject. Unfortunately, our paper was published only in Russian at that time, and it made no sense to give him the offprint. After returning to the US, Pines and C Herring published a note in *Physics Today* [8] reporting their meetings with Soviet scientists in which, in particular, a photo (Fig. 1) was published of one such meeting.

Together with Komberg, I decided to prove the weakness of the magnetic field in this binary pulsar based on a statistical analysis of the available pulsar data and using an anomalously small radio luminosity of pulsar with such a fast spin. Through a comparison with data on PSR 1916+16, we estimated its magnetic field to be $3 \times 10^{10}$ G, which was shortly thereafter confirmed by measurements [9]. We had time to publish our conclusions only as a preprint [10]. Before we submitted the paper, deceleration of the binary pulsar rotation was measured and its magnetic field was estimated. To a journal, we then submitted a revised version [11] of the preprint where all conclusions were made affirmatively and not conditionally, as in the preprint. Yet the prediction that the binary pulsar was old, was past the X-ray source stage, and had an anomalously weak magnetic field was published in my note in the Russian journal *Priroda* (*Nature*), together with the information on its discovery, before the deceleration of its rotation was measured.

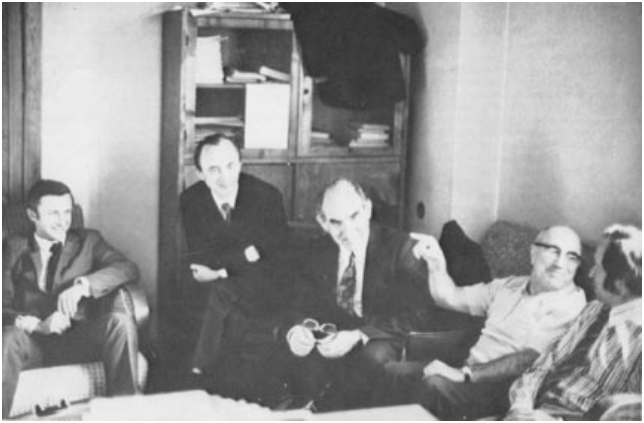

**Figure 1.** A meeting of theoreticians at the Landau Institute of Theoretical Physics. From left to right: G S Bisnovatyi-Kogan, I D Novikov, academicians V L Ginzburg and Ya B Zel'dovich, Professor David Pines. (Photo by G Baum from Ref. [8].)

Although all these ideas were published in 1974–1976, they became recognized only in the beginning of the 1980s after the discovery of millisecond pulsars. Nobody had doubts that recycled pulsars existed. The idea that the magnetic field of a neutron star can decay due to accretion was also widely recognized [13] after the discovery of the double pulsar system in which a recycled fast pulsar had a magnetic field more than two orders of magnitude smaller than a young slow pulsar [2]. The first binary pulsar proved to be a unique physical laboratory that allowed Taylor and his collaborators to carry out measurements confirming GR with a record accuracy. The Nobel Prize in physics was awarded to Taylor and Hulse, who discovered this pulsar through observations.

## 2. Magnetic fields and periods of radio pulsars

It was established already more than 35 years ago that radio pulsars are rapidly rotating strongly magnetized neutron stars in which the magnetic axis is inclined to the spin axis (see, e.g., [14]). A neutron star with an angular velocity $\Omega$ and a surface dipole magnetic field strength $B_p$ at its magnetic pole emits electromagnetic waves at the expense of its rotational energy losses $\dot{E}$ given by [15–17]

$$\dot{E} = -AB_p^2\Omega^4, \quad A = \frac{R^6 \sin^2\beta}{6c^3} = \frac{2d_m^2 \sin^2\beta}{3c^3 B_p^2}, \quad d_m = \frac{B_p R^3}{2}, \tag{1}$$

where $d_m$ is the magnetic moment of the star expressed through its radius $R$ and the magnetic field strength $B_p$ and $\beta$ is the angle between the magnetic and spin axes. The rotational energy of a homogeneously rotating star $E = I\Omega^2/2$ depends on its moment of inertia $I$. By measuring the spin period $P = 2\pi/\Omega$ and its decrease rate $\dot{P}$, one can estimate the magnetic field of the neutron star — a radio pulsar — as

$$B_p^2 = -\frac{I\dot{\Omega}}{A\Omega^3} = \frac{IP\dot{P}}{4\pi^2 A}. \tag{2}$$

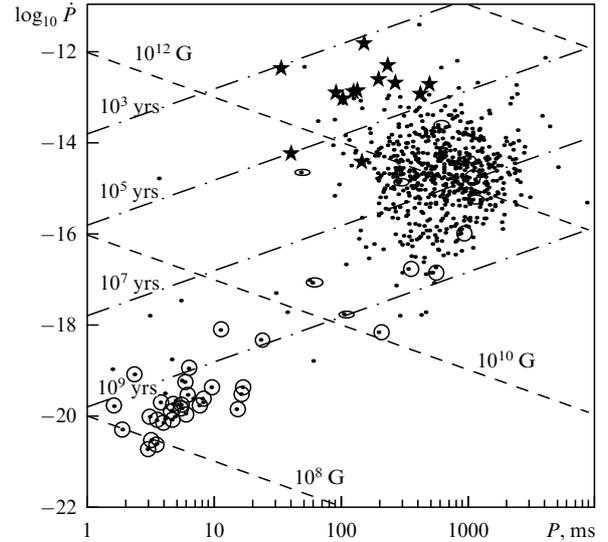

**Figure 2.** The location of pulsars on the $P - \dot{P}$ diagram (period–period derivative). Pulsars in binary systems with low-eccentricity orbits are encircled, and in high-eccentricity orbits are marked with ellipses. Stars show pulsars suspected to be connected with supernova remnants [28].

It was noted in Ref. [18] that a rotating neutron star represents a unipolar electromagnetic generator, which, in the presence of surrounding plasma, produces a relativistic wind that leads to rotational energy losses even in the case of a co-axial rotator, which allows setting $\sin\beta = 1$ in the magnetic field estimate in Eqns (1) and (2). Magnetic fields of radio pulsars, as measured by their periods and period derivatives, fall into a broad range between $10^8$ and $10^{13}$ G (Fig. 2) [17, 19, 28]. There are two significantly different groups of neutron-star magnetic fields: fields of single pulsars that lie in the range $10^{11}$ to $10^{13}$ G, and fields of recycled pulsars (RPs) occupying the range $10^8$ to $10^{10}$ G. The recycled pulsars are members of binary systems or were in such systems earlier. Single radio pulsars gradually lose their rotational energy and their periods are observed substantially above the value 0.033 s for the youngest pulsar in the Crab nebula. RPs have accelerated their rotation at the preceding accretion stage, and therefore rotate much faster, with periods as small as 1.558 ms in the case of PSR B1937+21. Significant magnetic field decay also occurs at the accretion stage, possibly due to screening by the accreting material. The prediction of the existence of rapidly spinning pulsars with a relatively weak magnetic field was made in Ref. [6] published before the first RP B1913+16 with the period 0.059 s was discovered [1].

For almost seven years after the discovery of pulsars, only single objects were found. This made the impression that pulsars avoid binary systems, although at least half of all ordinary stars enter binary systems. Over these seven years, it was assumed that the core-collapse supernova explosion, forming a pulsar, leads to the disruption of the binary system, or for some reason core-collapse supernovae completely avoid exploding in close binaries [20].

## 3. Binary X-ray pulsars

Observations of low-mass X-ray binaries (LMXBs) provide evidence for small magnetic fields of neutron stars that do not show X-ray pulsations typical of X-ray pulsars. Sig-nificant improvement in the accuracy of X-ray measurements by the RXTE observatory led to the discovery of an X-ray millisecond pulsar in LMXB SAX J1808.8-3658 with the pulse frequency 401 Hz [21]. This discovery bridged the gap and confirmed the genetic relation of LMXBs to recycled pulsars predicted earlier [22]. Neutron stars in these sources share both rapid rotation and low magnetic fields, usually about $10^8$ G. By the beginning of 2005, six millisecond pulsars with the frequencies between 185 Hz and 599 Hz were discovered [23, 24]. These pulsars are strictly separated into two groups by their orbital periods. Three pulsars are in binaries with orbital periods from 2 to 4.3 hr and have brown dwarf companions with normal chemical composition. The other three pulsars form ultracompact binaries with orbital periods from 40 to 44 min and helium companions. The distinction between these two groups can be due to the different masses of the companions tidally captured by neutron stars. A star of a sufficiently high mass can have the hydrogen burning time smaller than the duration of the LMXB stage, ultimately transforming into a recycled pulsar with a helium white dwarf companion. If a sufficiently close pair is formed, the helium-degenerate star approaches the neutron star due to the loss of orbital angular momentum via gravitational wave emission; the accretion onto the neutron star can resume at much smaller orbital periods, giving rise to an LMXB of the second generation [25]. After the second mass transfer stage has begun, the orbital period increases again. The X-ray source 4U 1820-30 with the orbital period 11.4 min is apparently a relatively recently formed second-generation LMXB, while three-ms accreting pulsars, along with LMXB 4U 1626-67 showing a close orbital period ($P_{orb} = 41.4$ min), could have been formed $3 \times 10^8$ years ago. This estimate follows from calculations of the evolution of binary systems consisting of a neutron star and a helium white dwarf with gravitational-wave driven mass transfer [26].

Most LMXBs are transient sources (X-ray bursters). In addition to six X-ray pulsars demonstrating stable pulsations at all stages, 11 sources in LMXBs were found to show coherent pulsations only at the bursting stage with a pulsation frequency below 619 Hz.

## 4. Theoretical prediction. Unavoidable formation of binary recycled radio pulsars

The analysis of the further evolution scenario of binary X-ray pulsar Her X-1 in Ref. [6] showed that the companion to the neutron star in this binary system should become a white dwarf on a time scale much shorter than the cosmological one. After that, the accretion of matter onto the neutron star should stop and a radio pulsar should appear, provided that the neutron star keeps a rapid enough rotation and strong enough magnetic field. In paper [6], we showed that the neutron star rotation after the end of the accretion stage should remain sufficiently rapid. The equilibrium rotational period of an accreting neutron star, which is equal to the period of the accretion disk at the Alfven radius (where the magnetic pressure balances that of the plasma), is determined by the relation

$$P_{eq} \approx \frac{3 B_{12}^{6/7}}{L_{37}^{3/7}}, \qquad (3)$$

where $P_{eq}$ is measured in seconds, $B_{12} = B/10^{12}$ G, and $L_{37} = L/10^{37}$ erg s$^{-1}$. In the bight X-ray source Her X-1, the luminosity reaches $L_{37} \approx 1$ and the spin period is 1.24 s and decreases with the characteristic doubling frequency time $t_{ev} \sim 3 \times 10^5$ years [27]. During its evolution, the radius of the optical star increases, which under the condition of filling its Roche lobe leads to an increase of the mass transfer rate $\dot{M}$, as does the X-ray luminosity, which is directly proportional to the mass transfer rate $L \approx 0.1 \dot{M} c^2$. Thus, the neutron star in Her X-1 must accelerate its rotation with time. When becoming a white dwarf, the star does not suffer an explosive mass loss, and hence a close binary with approximately the same binary parameters should form. The transition to the white dwarf stage and the accretion cessation occur on a time scale much shorter than $t_{ev}$, and therefore, after the accretion has stopped, the neutron star spin period should be much shorter than one second. Many radio pulsars spin much more slowly, and over the galactic lifetime, many systems like Her X-1 should be transformed into close binaries consisting of a rapidly rotating neutron star and a white dwarf companion.

The question arose as to why such close binary radio pulsars had not been discovered over the course of seven years? The only possible answer suggested in Ref. [6] was that the anomalous weakness of recycled radio pulsars in binaries was due to a substantial reduction of the magnetic field at the accretion stage. The field decay mechanism was assumed to be due to field screening by plasma accreted onto the neutron star over the long-term accretion stage. For the neutron star in Her X-1, the magnetic field estimate in (3) yields the upper limit $B < 3 \times 10^{11}$ G, while the estimate based on the rotation deceleration suggests $B > 10^8$ G. At lower fields, the X-ray pulsar stops pulsating because the magnetospheric radius of the neutron star becomes smaller than its physical radius. In [6], we assumed that the magnetic field decreases by 100 times due to accretion, which makes the newborn recycled pulsar shine more dimly than ordinary single pulsars with the fields $10^{11} - 10^{13}$ G do.

The discovery of the first binary radio pulsar confirmed this hypothesis. Pulsar PSR 1913 + 16 with the spin period 0.059 s turned out to be old, with the age $\sim 10^8$ years and magnetic field $2.3 \times 10^{10}$ G. Other recycled pulsars, whose number is already more than a hundred, share similar properties [2, 28]. Most such pulsars have low-mass normal or helium companions. These pulsars, originated from LMXBs, often show very rapid rotation, reaching millisecond periods, and their magnetic fields approach $10^8$ G. A smaller fraction of the recycled pulsars enter binary systems with a massive companion, which is another neutron star. Only in one such system recently discovered are both neutron stars observed as pulsars, but only one of them is recycled.

Periods of recycled pulsars with massive companions are usually above 20 ms, i.e., an order of magnitude higher than those of RPs with low-mass companions. This can be related to the evolutionary time of the companion during which the recycling occurs. A low-mass companion can provide matter for a very long time, almost to complete exhaustion ($10^9 - 10^{10}$ years), which is sufficient to accelerate the neutron star rotation to millisecond periods. Evolution of a massive companion, which ends in a supernova explosion, takes a much shorter time, of the order $10^6 - 10^7$ years, during which the neutron star is unable to strongly accelerate its rotation and the magnetic field cannot be strongly screened and stays above $10^{10}$ G. Binaries consisting of two neutron stars usually avoid globular clusters, which indicates their origin from primordial pairs of sufficiently massive stars.

Present estimates of the magnetic field in Her X-1 are quite controversial. As noted above, estimates [6] based on measurements of the luminosity, period, and its derivative yielded $10^{10}$ G. In 1978, the hard X-ray spectrum of this source was measured and a spectral feature at the energy $\sim 60$ keV was discovered [29]. This feature was interpreted as a cyclotron line corresponding to the magnetic field $\sim 6 \times 10^{12}$ G. Subsequent measurements showed that the spectral position of this feature varied in the range from 40 to 60 keV (see [30]). Such a strong discrepancy in the magnetic field estimate of Her X-1 could not be simply explained, and for a long time, the magnetic field of Her X-1 was accepted to be high, in agreement with the cyclotron-line energy. Detailed measurements of the X-ray pulse profile of Her X-1 first done by the ASTRON satellite [31] showed that the best-fit model requires a weak magnetic field of the order of the original estimate. These results were confirmed by measurements by other satellites [32, 33]. The location of the large energy cyclotron line has been reconciled with the magnetic field strength much smaller than its cyclotron value in a model in which the line is emitted by strongly anisotropic ultrarelativistic electrons oscillating along the magnetic field and rotating nonrelativistically in the perpendicular plane [30]. The line emission by such electrons, called magnetodipole, is shifted toward hard energies by the factor $2\gamma$, where $\gamma$ is the longitudinal relativistic factor. If $2\gamma \sim 40$, the broad emission line at the energy $\sim 50$ keV can emerge in the source spectrum at $B \sim 5 \times 10^{10}$ G.

A statistical analysis of 24 binary radio pulsars in almost circular orbits carried out in paper [34] revealed correlations between the neutron star spin period $P_p$ and the orbital period $P_{oeb}$, as well as between the orbital period and the magnetic field $B$. For binary radio pulsars with large orbital periods $P_{orb} > 100$ days, $P_p$ and $B$ increase with $P_{orb}$, and for smaller orbital periods, there is a dispersion around $P_p \sim 3$ ms and $B \sim 2 \times 10^8$ G. These correlations suggest that the increase in the accreting mass leads to a decrease in the magnetic field down to a bottom value of $10^8$ G.

## 5. Magnetic field decay during accretion

The first evidence for accretion-induced magnetic field decay was obtained from observations of the first binary pulsar PSR 1913+16 [1]. This pulsar rapidly rotates with the period 0.059 s in an orbit with the period $7^h 45^m$ and large eccentricity $\varepsilon = 0.6171$. Immediately after the discovery, this pulsar was interpreted to be an old recycled pulsar with a weak magnetic field [12]. The subsequent measurements [9] of the period derivative $\dot{P}$ allowed estimating its age as $\tau = P/2\dot{P} \approx 10^8$ years and magnetic field as $B \approx 2 \times 10^{10}$ G, which proved the proposed interpretation. In more than 100 recycled pulsars, the companion to the neutron star is usually a low-mass degenerate dwarf. Binaries consisting of two neutron stars, like PSR 1913+16, are much less numerous, with only six such systems known. All recycled pulsars, irrespective of the nature of their companions, have magnetic fields $10^8 - 10^{10}$ G, which fully confirms the hypothesis of the accretion-induced magnetic field decay. The discovery of the first binary system with two radio pulsars provided one more firm proof of this prediction.

According to simple estimates, there is a high probability of screening of the neutron star magnetic field by accretion. In Ref. [35], it was shown that if instabilities causing penetration of the accreting plasma inside the pulsar magnetosphere can be neglected, the pressure of the accreting plasma overcomes the magnetic pressure of a dipole with the strength $B = 10^{12}$ G in a time interval as short as one day, assuming the subcritical accretion rate $10^{-9}$ $M_\odot$ yr$^{-1}$. The original magnetic dipole then turns out to be buried under a plasma layer in which currents prevent the magnetic field of the star from emerging to the surface. In reality, instabilities lead to the penetration of gas through the magnetosphere and strongly slow down the process of magnetic field screening. During accretion, the magnetic field value is determined by the balance between the screening action of the accreting gas and turbulent field diffusion. As the accretion mass increases, the thickness of the layer through which the field should diffuse increases and, correspondingly, the external magnetic field decreases.

Many papers (see the references in [36]) considered the neutron star magnetic field decay due to the heating of the crystal crust of the neutron star during accretion that decreases the matter conductivity and accelerates the Ohmic dissipation of the magnetic field. This mechanism works, however, only if electric currents forming the pulsar magnetic field flow in the layers of the neutron star that can be heated by accretion. Clearly, the strongly degenerate neutron star core remains insensible to such a heating and preserves a very high conductivity. Therefore, the magnetic field formed by deep electric currents can be decreased only by the accreting plasma screening [6]. Several models of the accretion screening of the neutron star magnetic field have been studied [36–39]. It is usually assumed that the infalling matter is canalized by the magnetic field and flows along the field lines to the magnetic pole region of the dipole [40, 41]. As the magnetic field decays due to the screening, the canalization of the accretion flux decreases and the polar cap area increases. It is easy to estimate the polar cap angular width $\theta_P$ as a function of the neutron star surface magnetic field $B_s$ [42]. The magnetic force line that reaches the Alfvenic radius $r_A$ starts at the angle $\theta_P$ on the surface of a neutron star of radius $r_s$ and is the last closed line of the dipole field. It is easy to show that

$$\sin \theta_P = \left( \frac{r_s}{r_A} \right)^{1/2}. \quad (4)$$

Setting the dynamic pressure of freely falling gas equal to the magnetic pressure at the Alfvenic surface, we find the Alfvenic radius value

$$r_A = (2GM)^{-1/7} r_s^{12/7} B_s^{4/7} \dot{M}^{-2/7}, \quad (5)$$

where $M$ is the neutron star mass and $\dot{M}$ is the accretion rate. Equations (4) and (5) imply that

$$\sin \theta_P \propto B_s^{-2/7}. \quad (6)$$

Therefore, the polar cap size increases as the magnetic field decreases until $\theta_P$ becomes equal to $90°$ and Eqn (6) cannot be applied any more. For $M = 3 \times 10^{33}$ g, $\dot{M} = 10^{-8} M_\odot$ yr$^{-1}$, $r_s = 10$ km, $B_s = 10^{12}$ G, we obtain from Eqn (5) that $r_A \approx 260$ km, and from Eqn (4), we find the initial angular size of the polar cap of the order of $10°$. The accreting matter falls onto the polar caps and then spreads over the stellar surface toward the equator. In the model considered in

Refs [39, 43], the matter flows from both poles meet near the equator and go inside the star. In a kinematical model calculated in [39], the matter flows cover the star magnetic field and effectively decrease its strength; the sizes of the polar caps increase accordingly. When the polar cap size attains 90°, the accretion flow becomes spherical and radial and, as argued in [39], the magnetic field decay stops due to the assumed low efficiency of the magnetic field screening by the radial flow. If $\theta_{P,i}$ is the initial size of the polar cap, then according to Eqn (6) the magnetic field decreases by $(\sin 90°/\sin \theta_{P,i})^{7/2}$ times during the accretion stage. Assuming $\theta_{P,i}$ to be in the range $5°-10°$, we obtain the possibility of the accretion-induced magnetic field decay by $10^3-10^4$ times, which is exactly the difference between the magnetic fields of ordinary and recycled pulsars. Assuming that the magnetic field during accretion reaches some stationary value when the Alfvenic radius equals the neutron star radius, we obtain

$$B_{\text{asymp}} = (2GM)^{1/4} \dot{M}^{1/2} r_s^{-5/4} \quad (7)$$

from Eqn (5) with $r_A = r_s$. Using standard values of the parameters as discussed above we find $B_{\text{asymp}} \approx 10^8$ G. After the accretion has stopped, a recycled millisecond radio pulsar emerges with a magnetic field close to $10^8$ G. The development of magneto-hydrodynamic instabilities near the Alfvenic surface facilitates plasma penetrating the magnetosphere and slows down the process of the magnetic field screening by accretion, considered in Ref. [44].

It was noted in Ref. [11] that "neutron star magnetic fields screened by intensive accretion can percolate outwards after the accretion has stopped." In this picture, the neutron star magnetic field value should increase with time, which can increase the rotational energy loss rate $\dot{E}$ with time, in contrast to the usual decrease $\dot{E}$ with time. By writing the field growth as $B \sim \Omega^{-k}$, we obtain $\dot{E} \sim \Omega^{4-2k}$ in accordance with Eqn (1), and for $k > 2$, the rotational energy losses increase as the pulsar rotation decelerates. Another characteristic of the pulsar parameters is the braking index $n = \Omega \ddot{\Omega}/\dot{\Omega}$ (for the above relation, $n = 3 - 2k$). If observations yield $n < 3$, this may be due to the neutron star magnetic field increase with time [45]. The estimates of the magnetic field percolation outward were obtained in Ref. [46]. The authors considered the case of a very rapid percolation and the field growth over $10^3-10^4$ years until the original value before the accretion is recovered. Clearly, this conclusion contradicts observations because all recycled pulsars, including very old ones with the age up to $10^{10}$ years [28], have equally small magnetic fields. Such a discrepancy is due to the artificial model considered in Ref. [46]. The percolation in that model started from depths smaller than 260 m, which corresponded to the accreted mass smaller than $5 \times 10^{-5} M_\odot$. The conductivity is not very high in the outer layers of the neutron star, which allows rapid percolation of the field outward. Actually, the accretion stage can last up to $10^7 - 10^8$ years or more and the mass of the accreted matter can be several orders of magnitude higher. The magnetic field is then buried in much deeper layers of the star with much higher conductivity. Accordingly, in recycled pulsars, the magnetic field percolation outward becomes much less effective. As noted in Ref. [45], the magnetic field percolation outward can occur in young neutron stars, which accreted an insignificant amount of screening matter after the supernova explosion. In these stars, a sufficiently rapid percolation of the magnetic field and its growth on the surface is possible.

## 6. Evolutionary formation schemes

Evolutionary schemes of recycled pulsar formation were discussed in Ref. [6] and different scenarios of close binary evolution were considered in Ref. [47] (see Ref. [48] for a review). In Refs [6, 47], several scenarios for the evolution of sufficiently massive binary stars capable of producing neutron stars or black holes from one or two components were analyzed. In the course of stellar evolution, different stages of nuclear burning result in the formation of a white dwarf or, alternatively, quiet nuclear burning continues until iron-group elements are formed in the stellar core. After the formation of the iron core, the star collapses and forms a neutron star or a black hole. During the evolution of a close binary, intensive mass exchange between the binary companions can occur, and the initially more massive star can thus become less massive. A loss of stellar mass is also possible via matter outflow — the stellar wind. Several scenarios for the evolution of massive close binary systems are possible.

(1) The binary system can be destroyed after the evolution of the more massive companion is completed by the core collapse and the subsequent supernova explosion. The destruction of the binary can be due to either rapid ejection of a large amount of mass (the Blaauw effect) or an anisotropy of the explosion with anisotropic mass ejection or neutrino flux, or the complete destruction of the star in the explosion. The anisotropic explosion can be realized in the magneto-rotational model, which allows a mirror symmetry breaking of the magnetic field, leading to anisotropic mass ejection and neutrino flux [49, 50]. After the destruction of the binary and the neutron star formation in the first explosion, a radio pulsar moving at a high spatial speed is likely to appear. The companion of this binary, after completing its evolution, can also turn into a single radio pulsar. In this way, a spatially close pair of radio pulsars with comparable ages can emerge. Several candidates were proposed in Refs [6, 11]. The second collapse may also result in a black hole, which may be very difficult to observe. So far, no single black holes have been discovered.

(2) If the close binary survives after the first collapse with a possible supernova explosion, the relativistic object (neutron star or black hole) becomes an X-ray source when the accretion of mass from the secondary companion begins. During the accretion stage, the neutron star rotation accelerates and the magnetic field decreases. If the second collapse and explosion unbinds the binary, a spatially close pair of pulsars, one of which is recycled, can be formed. So far, no such pairs have been found. Because the magnetic axes of pulsars can become nonparallel in an asymmetric explosion, both genetically related pulsars are not necessarily seen from the earth because of their narrow beams. The destruction of the binary during the formation of a black hole appears less likely than during the formation of a neutron star. This, however, seems possible, for example, due to neutrino flux asymmetry [50].

(3) The discovery of low-mass X-ray binaries demonstrated that, undoubtedly, such a binary is not destroyed after the end of the evolution of the primary companion and a recycled pulsar in a close binary system is formed [6]. The companions of LMXBs are low-mass stars such that they end their evolution as white dwarfs. The companions of most recycled pulsars in close binaries are low-mass white dwarfs.

(4) The formation of a close binary comprising a neutron star and a high-mass companion that would collapse and

explode to form another neutron star occurs much less frequently; close neutron-star binaries constitute about 5% of the number of recycled pulsars, although the first recycled pulsar discovered belongs to this type. These pairs avoid globular clusters and their evolution is not affected by collisions with surrounding stars [28, 51].

Analysis of the reasons for the absence of pulsars in close binaries carried out by us in 1973 [6] led to the following conclusions:

"... evolutionary considerations do not just allow the possibility of the existence of a neutron star in pair with another star, which is also at the advanced evolutionary stage (white dwarf, neutron star, collapsar), but, moreover, make such a conclusion very probable."

The transition of the companion to a dead star (the white dwarf in Her X-1 or a neutron star) is accompanied by an a decrease of the accretion rate and an increase of equilibrium period (3). According to the estimates in [6], the drop in the accretion rate occurs so rapidly that there is no time for equilibrium period (3) to establish and after the completion of accretion, a 'naked' rapidly rotating neutron star should be left. We concluded in Ref. [6] that "the reason for the absence of radio pulsars in close binary systems is the neutron star magnetic field decay as it accretes matter from the central component."

These conclusions were fully confirmed by later observations.

## 7. Binary radio pulsars in globular clusters. The formation of single recycled pulsars

The number of RPs is rapidly growing due to dedicated searches. This especially relates to RPs in globular clusters (GCs). In 1990, 10 RPs were known, their number increased to 30 in 1992, and about 100 RPs had been discovered by the end of 2004 [52]. The classes of RPs are divided in two unequal parts: 5% belong to relativistic recycled pulsars (RRPs) consisting of two neutron stars, and all other RPs include binaries with low-mass degenerate (white) dwarf companions and a single RP. The period of all RRPs is above 23 ms, while many other RPs have millisecond periods. RRPs avoid GCs and are the end product of the evolution of sufficiently massive close binaries. They represent a unique tool to study properties of gravity, and future observations of RP J0737−3039A,B will significantly increase the accuracy of measurements of GR effects and possible deviations from GR.

Many pulsars considered to belong to the recycled class (rapid rotation + weak magnetic field) are solitary. The disruption mechanism of a binary to which the recycled pulsar had belonged is not yet fully understood. After the first eclipsing radio pulsar had been discovered, the hypothesis was put forward that the companion can be evaporated by pulsar radiation [53], but RP statistics do not support this. RPs, as well as LMXBs, are concentrated in GCs. According to [28], 47 of the total 103 RPs enter GCs, while the mass of all GCs does not exceed 0.001 mass of the galaxy. Thus, the relative concentration of RPs in GCs is almost a thousand times as large as the average galactic one. A similar concentration of LMXBs in GCs is observed [54]. Such a similarity in RP and LMXB distributions provides another clue to their generic relation [6, 11, 22]. According to [28], 21 single RPs belong to GCs (45% of all RPs in GCs), while only 9 single RPs are found in the galactic bulge, which amounts to only 16% of their total number in the bulge. Such a strong difference indicates that not only the formation of LMXBs and their subsequent transition to RPs are related to their entering GCs [55], but the disruption of binaries containing RPs can also be due to their frequent encounters with GC stars, where the stellar concentration is much higher than in the bulge. It is in Ref. [25] assumed that single RPs in the bulge can be the remnants of fully evaporated GCs [55], as well as bulge LMXBs [55] that later transform into bulge RPs. Clearly, the evaporation time of GCs in which single RPs form must be sufficient for two evolutionary stages: LMXB formation by tidal capture or stellar exchange in an encounter with an initially close binary [56], and then RP formation due to LMXB evolution or the disruption of the binary and the formation of a single RP. Numerical simulations [57] showed that collisions with GC stars are capable of disrupting only quite wide binaries with orbital periods of 10–100 days. However, observations suggest [28] that RPs usually reside in much closer binaries. The solution to this problem can be found in the framework of the 'induced evaporation' (IE) mechanism proposed in Ref. [26], according to which the disruption of the most compact binaries occurs at the late stages of the LMXB evolution.

It was theoretically predicted as early as 1950 [58] that stellar collisions with close binaries are accompanied by energy extraction from the binary system and the heating of the entire cluster. In contrast to this, stellar collisions with a quite wide binary lead to its further widening, ending up with total disruption. These results where confirmed by numerical experiments [59], which, in particular, implied that the formation of several close binaries can lead to evaporation and total disruption of the GC. This property has a simple physical explanation. The energy exchange during an elastic collision of stars is determined by their initial kinetic energies (velocities for stars of equal masses). On average, the energy is passed from the more rapidly moving to the more slowly moving star, and this property is independent of whether the star is single or enters a binary. Collisions between single stars result in a quasi-stationary velocity distribution close to the Maxwellian one, with the deviations at high velocities being due to evaporation from the cluster. Collisions with a close binary system result in the following. In a close binary, the orbital kinetic energy of the components exceeds the mean kinetic energy of the stars in the cluster, and hence, on overage, a star from the cluster acquires energy in the collision, decreasing the total energy of the binary. The absolute value of the total negative energy of the binary (kinetic + potential) becomes larger, and therefore, in accordance with the virial theorem [15], the orbital kinetic energy increases. The binary becomes more compact and transfers energy to the surrounding stars more efficiently, thus heating the cluster. In other words, we can say that the binary system has a negative heat capacity because the loss of its energy increases the kinetic energy of the stars, which is the analog of temperature. A sufficiently effective energy exchange between the cluster stars and close binaries can lead to cluster disruption in a time scale shorter than the cosmological one, leaving behind LMXBs and binary and single RPs as remnants that become bulge objects.

Another situation occurs during collisions of field stars with a close binary system in which one of the components is a low-mass degenerate star that fills its Roche lobe and transfers matter to the more massive neutron star. Let such a binary consist of a neutron star with mass $M_1$ and a degenerate dwarf of mass $M_d$. If the degenerate dwarf fills its Roche lobe, its radius is related to the distance between the components $R_{12}$ by

$$R_d = R_{12} \frac{2}{3^{4/3}} \left(\frac{M_d}{M}\right)^{1/3}, \quad M = M_1 + M_d. \quad (8)$$

Equilibrium models of low-mass degenerate dwarfs with different chemical compositions with nonideal matter were calculated in Ref. [60]. The radius of a star with the equation of state of an ideal nonrelativistic degenerate gas increases with decreasing mass until nonideal effects leading to repulsion between particles change this dependence. This occurs when the mass of the dwarf decreases to several masses of Jupiter. The angular momentum of the binary is lost due to gravitational radiation [61] and the system becomes more compact. When the companion approaches the neutron star such that it fills its Roche lobe, the mass transfer onto the neutron star begins. During the mass transfer, the radius of the companion is close to its Roche lobe, and hence decreasing the mass of the degenerate dwarf is accompanied by an increase in both its radius and the Roche lobe radius. This means that the binary separation increases and the system becomes softer. The evolution of such binaries driven by gravitational radiation until the white dwarf matter becomes nonideal is calculated in Ref. [26]. It is found that over the cosmological time $\tau_c = 2 \times 10^{10}$ years, the mass of a carbon white dwarf decreases to $0.0025 M_\odot$. By that time, the orbital period of the binary system $P_{\rm orb}$ is 1.5 h,

$$P = \frac{2\pi R_{12}^{3/2}}{(GM)^{1/2}} = \frac{9\pi R_d^{3/2}}{\sqrt{2GM_d}}, \quad (9)$$

and the orbital velocity of the dwarf in the binary is

$$v_d = \frac{2\pi R_{12}}{P} = \frac{\sqrt{2}}{3^{2/3}} \left(\frac{GM^{2/3}M_d^{1/3}}{R_d}\right)^{1/2} \text{ for } M_d \ll M. \quad (10)$$

The binary becomes 'soft' and is disrupted by close encounters with field stars if the orbital kinetic energy of the dwarf is smaller than the mean kinetic energy of the field stars, $M_f v_f^2 > M_d v_d^2$. For globular clusters 47 Tuc and $\omega$ Cen, the corresponding velocities are 10.3 and 16.8 km s$^{-1}$, and the respective mean masses are $0.67 M_\odot$ and $0.51 M_\odot$. According to the estimates in [26], binaries in these GCs are soft if the respective masses of the dwarfs are below $6 \times 10^{-4} M_\odot$ and $9 \times 10^{-4} M_\odot$, which is several times smaller than the lowest masses the dwarfs can attain in the gravitational-radiation-driven mass transfer over cosmological time. To reach these masses, 10 cosmological times would be required.

In Ref. [26], we drew attention to the fact that collisions of the field stars with close binaries, in which a white dwarf fills its Roche lobe and transfers matter to the neutron star, have the same effect on the binary evolution as gravitational radiation does. A hard binary loses energy and angular momentum in collisions with the field stars, and in the absence of mass transfer it would become even harder; but if mass transfer from the Roche-filling degenerate dwarf occurs, collisions make the binary softer under all conditions. For a sufficiently large concentration of the surrounding stars, the collision-induced evolution of the binary can occur faster than it can due to gravitational radiation. A close binary becomes softer and is directly disrupted by close stellar encounters. This process of binary disruption was called 'induced evaporation' [26]. To estimate the characteristic induced evaporation time $\tau_m$, we use the relaxation time for momentum [62], because in a close binary system, the excessive momentum and energy can be transferred to the barycenter motion:

$$\tau_m = \frac{v_d^3}{4\pi G^2 M_f^2 n_f \Lambda}. \quad (11)$$

Here, the Coulomb logarithm $\Lambda \approx 10$ and $n_f$ is the mean concentration of the field stars. Using Eqn (10), we can write the characteristic relaxation time determining the disruption time due to IE as

$$\tau_m \approx 10^{22} \frac{m m_d}{m_f^2 n_{51}}, \quad (12)$$

where $\tau_m$ is in seconds, $m$ is the mass of the star in solar units, and $n_{51} = n_f / 10^{-51}$ cm$^{-3}$. Here, we have used the approximate mass–radius relation for white dwarfs $R = 7.6 \times 10^8 m_d^{1/3}$, which is valid for the equation of state of an ideal electron Fermi gas. Everywhere, we considered a carbon white dwarf with the number of baryons per electron $\mu_e = 2$. By examining the evolution of a binary due to collisions, we showed in Ref. [26] that to disrupt a binary by the IE mechanism on a time scale shorter than the cosmological one, the concentration of the field stars $n_f$ (pc$^{-3}$) must be higher than

$$n_f > 3 \times 10^5 \frac{m^{9/11}}{m_f^2}. \quad (13)$$

The central parts of the most dense GCs like M15, where many single RPs have been discovered, satisfy this requirement. Therefore, the IE mechanism can lead to the formation of single RPs in GCs. RPs in the bulge can result from evaporation of GCs in which they have already formed. We note that binaries with hydrogen–helium brown dwarfs are disrupted by the IE mechanism more efficiently [26], at a smaller star concentration; in Eqn (13), the factor 2 should be used instead of 3.

## 8. Relativistic effects in compact neutron star binaries

### 8.1 Observed relativistic effects
Because neutron stars are very compact massive objects, a double pulsar and other binaries comprising two neutron stars can be regarded as almost ideal point-like objects and can be used to test the theory of gravity in the strong-field limit. To perform such a testing observationally, different relativistic corrections to Kepler's orbit determination should be found, which are called post-Keplerian (PK) parameters. For point-like objects with insignificant stellar rotation effects, PK parameters in any theory are functions of the masses of both stars, which are unknown beforehand, and of the Keplerian parameters, which can be measured with high accuracy: $P_b$ (the orbital period) and $e$ (the eccentricity).

When only the masses of stars are free parameters, measurements of three or more PK parameters make the system of equations overdetermined and, hence, allow testing theories of gravity. When the correct theory of gravity is used, all curves on the mass diagram of two neutron stars $M_A$ and $M_B$, determined by the measured PK parameters, should intersect at one point irrespective of the theory. In GR, the five most important PK parameters in the first post-Newtonian [1PN, or $\mathcal{O}(v^2/c^2)$] order are given by [63]

$$\dot{\omega} = 3 T_\odot^{2/3} \left(\frac{P_b}{2\pi}\right)^{-5/3} \frac{1}{1-e^2} (M_A + M_B)^{2/3}, \quad (14)$$

$$\gamma = T_\odot^{2/3} \left(\frac{P_b}{2\pi}\right)^{1/3} e \frac{M_B(M_A + 2M_B)}{(M_A + M_B)^{4/3}}, \quad (15)$$

$$\dot{P}_b = -\frac{192\pi}{5} T_\odot^{5/3} \left(\frac{P_b}{2\pi}\right)^{-5/3} \frac{1}{(1-e^2)^{7/2}}$$
$$\times \left(1 + \frac{73}{24} e^2 + \frac{37}{96} e^4\right) \frac{M_A M_B}{(M_A + M_B)^{1/3}}, \quad (16)$$

$$r = T_\odot M_B, \quad (17)$$

$$s = x T_\odot^{-1/3} \left(\frac{P_b}{2\pi}\right)^{-2/3} \frac{(M_A + M_B)^{2/3}}{M_B}, \quad (18)$$

where $P_b$, $e$, and $x$ are the period, eccentricity, and major semiaxis of the binary orbit derived from observations. The masses $M_A$ and $M_B$ relate to pulsars A and B in the double pulsar J0737−3039 or to the pulsar and its invisible companion in other neutron star binaries. Both masses are expressed in solar units ($M_\odot$). The constant $T_\odot$ is defined as $T_\odot = GM_\odot/c^3 = 4.925490947$ µs, where $G$ and $c$ are the Newton gravity constant and the speed of light. The first PK parameter, $\dot{\omega}$, is measured most straightforwardly and gives the rate of relativistic periastron (apse line) advance of the orbit. According to Eqn (14), the measurement of $\dot{\omega}$ immediately yields the total mass of the system $(M_A + M_B)$ [64]. The parameter $\gamma$ represents the amplitude of the signal time delay due to variable gravitational redshift and time dilation (quadratic Doppler effect) when the pulsar moves in an elliptical orbit, caused by the variable distance between the stars and the pulsar velocity. Emission of gravitational waves results in the loss of the orbital angular momentum and decreases the orbital period ($\dot{P}_b$). The remaining two parameters $r$ and $s$ determine the time delay due to the Shapiro effect and are related to the companion's gravitational field. These parameters can be inferred from sufficiently accurate timing observations only if the binary orbit is observed almost edge-on.

### 8.2 Relativistic effects in the Hulse–Taylor pulsar
The orbital period of the pulsar PSR 1913 + 16 is $P_b \approx 7.8$ h, the eccentricity is $e \approx 0.61$, and the mass function is $f_A \approx 0.13 M_\odot$. The mass function is derived from Kepler's laws as [65]

$$f_A(M) = \frac{M_B^3 \sin^3 i}{(M_A + M_B)^2} = 10385 \times 10^{-11} (1-e)^{3/2} K_A^3 P_b, \quad (19)$$

were the masses are in solar units, the orbital period is in days, the semi-amplitude of the radial velocity curve of the pulsar $K_A$ is in km s$^{-1}$, and $i$ is the binary inclination angle, $i = 90°$ for the edge-on orbit. Observed values enter the right-hand side of this formula, and therefore the mass function can be calculated from observations. Table 1 lists the pulsar parameters with 1σ uncertainties, including three relativistic parameters $\langle\dot{\omega}\rangle$, $\gamma$, and $\dot{P}_b$ obtained by fitting pulsar timing measurements from 1981 to 2001 [66]. The level of accuracy achieved is so high that small kinematical corrections (about 0.5% of the observed $P_b$), caused by the acceleration of the solar system and the binary system with the recycled pulsar in the gravitational field of the galaxy [67], must be taken into account. Accounting for all these factors allows one to conclude that all parameters agree with GR within the accuracy not worse than 0.4% accuracy (see Fig. 3).

**Table 1.** Measured orbital parameters of PSR 1913 + 16 [66].

| Parameter | Value | Uncertainty |
|---|---|---|
| $a \sin i/c$, s | 2.341774 | 0.000001 |
| $e$ | 0.6171338 | 0.0000004 |
| $P_b$, d | 0.322997462727 | 0.000000000005 |
| $\omega_0$, deg | 226.57518 | 0.00004 |
| $\langle\dot{\omega}\rangle$, deg yr$^{-1}$ | 4.226607 | 0.000007 |
| $\gamma$, c | 0.004294 | 0.000001 |
| $\dot{P}_b$, $10^{-12}$ s s$^{-1}$ | −2.4211 | 0.0014 |

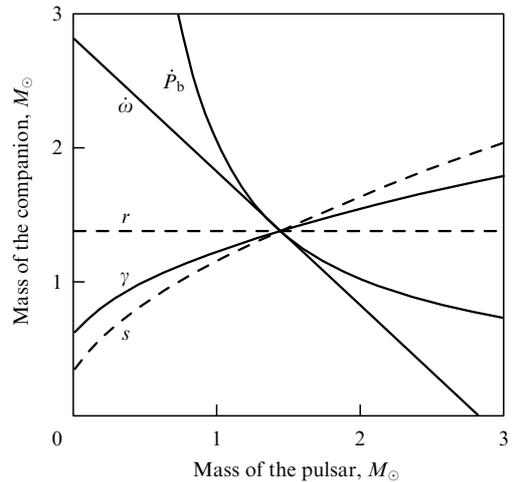

**Figure 3.** Observational constraints on the component masses in the binary pulsar PSR 1913 + 16. The solid curves correspond to Eqns (14)–(16) with measured values of $\dot{\omega}$, $\gamma$, and $\dot{P}_b$. The intersection of these curves at one point (within an experimental uncertainty of about 0.35% in $\dot{P}_b$) proves the existence of gravitational waves. The dashed lines correspond to the predicted values of the parameters $r$ and $s$. These values can be measured by future accumulation of observational data [67].

## 9. Two pulsars in a binary system: the most compact neutron star binary as the best laboratory for testing GR

### 9.1 Pulse arrival time in the binary pulsar
The discovery of RP J0737-3039A with the period 23 ms around another compact object with the orbital period 2.4 h in a low-eccentricity orbit ($e = 0.0878$) was reported in Ref. [68]. The binary parameters suggested another neutron star as the companion. In Ref. [2], this companion was reported to be the radio pulsar J0737f-3039B with the spin period 2.8 s. Thus, for the first time, a binary system consisting of two pulsars, a recycled one and a normal one, was discovered. This binary has the shortest orbital period known among all RPs with two neutron stars. Two pulsars with quite narrow beams in a short-period binary yield unprecedented opportunities to test fundamental laws of gravitational physics and make this system a remarkable laboratory for relativistic astrophysics research. Over one year, all five relativistic parameters (14)–(18) were measured for pulsar A. In addition to checking the PK parameters, new possibilities to test GR (as compared to the Hulse–Taylor pulsar) open up here because the orbital parameters of both pulsars can be measured independently. Because pulsar A rotates much faster and shines more brightly all the time except for a $\sim 27$ second eclipse, the arrival times of pulses from pulsar A are measured much more accurately than from pulsar B, and hence the PK parameters of the pulsar A orbit are determined with much higher precision. By measuring projections of major semiaxes of orbits of both pulsars A ($x_A$) and B ($x_B$), the precise value of the mass ratio $R(M_A, M_B)$ of both stars can be found from Kelper's third law:

$$R(M_A, M_B) \equiv \frac{M_A}{M_B} = \frac{x_B}{x_A}. \quad (20)$$

In the first PN order at least, this simple relation for $R$ is valid in any relativistic theory of gravity. More importantly, the value of $R$ is independent of both theory and strong self-gravity effects affecting the PK parameters. This allows obtaining a new strong constraint on the theory of gravity, because any combination of masses determined from measurements of PK parameters must correspond to the mass ratio as derived from Kepler's third law. Combined with the five known PK parameters, this additional constraint available in the double-pulsar system enables finding the most overdetermined (by the present time) system of equations, in which most effects can be studied in the strong-field approximation. Due to the emission of gravitational waves, the distance between pulsars in this system decreases at a rate of 7 mm per day. The unique mass ratio valid in both GR and any other theory of gravity can be utilized in plotting all PK parameters on the component mass diagram $M_A - M_B$. If the correct theory of gravity is used (GR in this case), all these curves must intersect the given line $R$ = const at one point. Up to now, such tests have been available only for PSR 1913 + 16 [66] and PSR B1534 + 12 [70]. However, for these systems, it has been impossible to present equally many curves on the $M_A - M_B$ plane as for the double pulsar, shown in Fig. 4 taken from Ref. [71].

By another lucky chance, the double pulsar system is observed almost edge-on, which has allowed high-precision measurements of the Shapiro delay. This geometry has allowed investigating the interacting pulsars' magnetospheres [72], the eclipse of pulsar A, and other effects of this interaction [2, 73, 74]. These effects should be taken into account in analyzing times of arrival (TOAs) of pulses when the assumption of 'purely' gravitating point masses can be broken. So far, all studies have confirmed the 'pure' nature of this system, but it should be borne in mind that they have been carried out by assuming the constant average pulse shape with a high signal-to-noise ratio [75]. Changes of the pulse shape could lead to systematic changes in measured TOAs, and therefore a detailed analysis of pulse profiles from pulsars A and B is needed in order to find any changes of the pulse shapes with time. There are several reasons for such changes in the double-pulsar system.

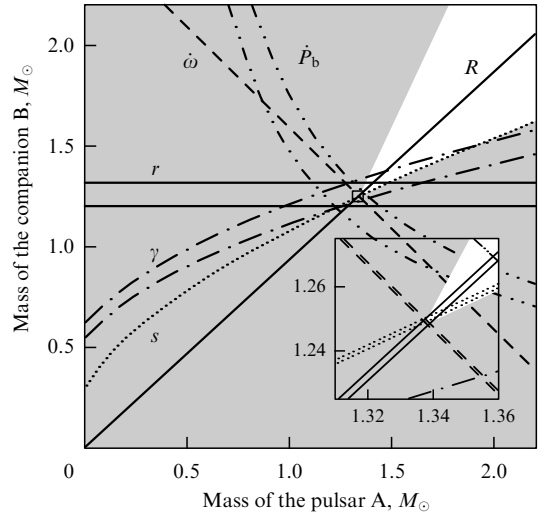

**Figure 4.** Observational constraints on the neutron star masses in the double pulsar J0737-3039 on the $M_A - M_B$ plane. The shadowed region show the values prohibited by the mass functions of the companions (19). Other bounds are presented by pairs of lines bordering masses allowed by GR as inferred from the corresponding PK parameters ($\dot\omega$, $\gamma$, $\dot P_b$, $r$, and $s$) with the known mass ratio (the line $R$). The right box inset shows the area around the intersection of the three strongest restrictions ($\dot\omega$, $R$, $s$), magnified 16 times. The allowed region lies at the intersection of these three bands.

In GR, the rest frame of a freely falling object relative to a remote observer precesses (the so-called geodetic precession) due to the spin–orbit interaction, similar to the spin–orbit interaction in atomic physics [76]. The pulsar's spin precesses around the total angular momentum vector, thereby changing the orientation of pulsars with respect to each other and the earth. Because the orbital angular momentum is much larger than the proper angular momenta of neutron stars, the total angular momentum virtually coincides with the orbital one. The precession rate [77] depends on the orbital period and eccentricity, as well as on the masses of pulsars $M_A$ and $M_B$. For the orbital period of the double pulsar, the geodetic precession periods calculated in GR are 75 years and 71 years for pulsars A and B, respectively.

Geodetic precession directly affects the TOAs because it changes the spin directions of the stars and hence alters the aberration effects [69]. These changes alter the 'observed' values of the projection of the major semiaxis and eccentricity, which differ from the 'intrinsic' values due to variable aberration, which potentially allows more accurate determination of the system's geometry [78]. The geodetic precession also results in secular changes in the pulse profiles from both pulsars due to the pulsar beam precession together with the spin axis, as well as due to the change of the angle between the spin and pulsar axes.

### 9.2 Results of data analysis [71]
In 15 months of observations, no changes in the mean pulse profile from pulsar A were detected. This strongly facilitates the analysis of TOAs from pulsar A. However, such changes may be discovered in the future. The smallness of the geodetic precession effects is probably due to the pulsar A spin being almost parallel to the orbital angular momentum, but it cannot be excluded that the system being at the precession phase, which is difficult to observe, is responsible for this smallness. It is this situation that occurs for PSR B1913 + 16

**Table 2.** Observed and calculated parameters of pulsars PSR J0737-3039A and B. Standard errors (1σ) in the last digit are shown in parentheses [2, 71].

| Parameter | Pulsar | |
|---|---|---|
| | PSR J0737-3039A | PSR J0737-3039B |
| Pulsar period $P$, ms | 22.699378556138(2) | 2773.4607474(4) |
| Period derivative $\dot{P}$ | $1.7596(2) \times 10^{-18}$ | $0.88(13) \times 10^{-15}$ |
| Right ascension $\alpha$ (J2000) | $07^{\rm h} 37^{\rm m} 51^{\rm s}.24795(2)$ | |
| Right ascension $\delta$ (J2000) | $-30°39'40''.7247(6)$ | |
| Dispersion measure DM, cm$^{-3}$ | 48.914(2) | 48.7(2) |
| Orbital period $P_{\rm b}$, d | 0.1022515628(2) | |
| Eccentricity $e$ | 0.087778(2) | |
| Periastron advance rate $\dot{\omega}$, deg yr$^{-1}$ | 16.900(2) | |
| Large semi-major axis projection $x = a \sin i/c$, s | 1.415032(2) | 1.513(4) |
| Gravitational redshift parameter $\gamma$, ms | 0.39(2) | |
| Shapiro delay parameter $s = \sin i$ | 0.9995(4) | |
| Shapiro delay parameter $r$, μs | 6.2(6) | |
| Orbital period decay rate $\dot{P}_{\rm b}$, $10^{-12}$ | $-1.20(8)$ | |
| Mass ratio $R = M_{\rm A}/M_{\rm B}$ | 1.071(1) | |
| Radio flux density at 1390 MHz, mJy | 1.6(3) | 0 – 1.3(3) |
| Characteristic age $\tau$, $10^6$ yrs | 210 | 50 |
| Surface magnetic field $B$, G | $6.3 \times 10^9$ | $1.6 \times 10^{12}$ |
| Rotational energy loss rate $\dot{E}$, erg s$^{-1}$ | $5800 \times 10^{30}$ | $1.6 \times 10^{30}$ |
| Mass function, $M_\odot$ | 0.29097(1) | 0.356(3) |
| Distance, kpc | $\sim 0.6$ | |

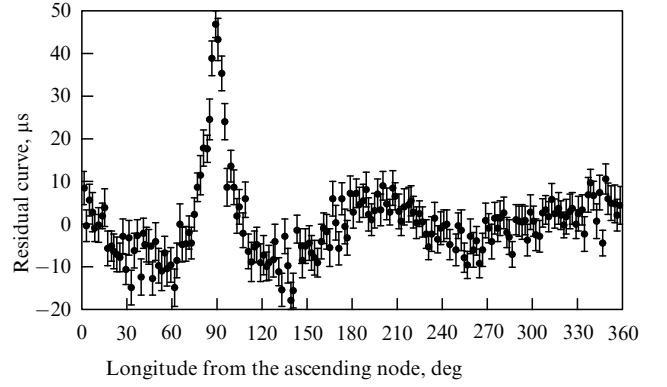

**Figure 5.** The Shapiro delay caused by the gravitational field of pulsar B in measuring TOA from pulsar A. Shown is the residual curve that is obtained after including all PK effects listed in Table 2 except the Shapiro delay parameters $r$ and $s$. The feature to the left is due to the strongest harmonics of the Shapiro delay that remains after the other PK parameters are taken into account.

[79]. A quite wide range of system geometry parameters remains possible, but some models predicting strong pulse profile change can already be rejected [80].

In contrast to pulsar A, the study of the profile change in pulsar B has revealed significant secular variation in addition to the orbital one. The orbital light curve of pulsar B also changes. This can be due to the interaction between two pulsar magnetospheres caused by the geometry change during geodetic precession. Accounting for these variations significantly complicates the TOA data analysis. Preliminary results of this analysis are listed in Table 2. The results imply a surprisingly small proper velocity of the system (30 km h$^{-1}$), considering the dispersion measure that suggests a distance of 600 pc [81]. Such a slow proper motion can be due to a small recoil in the supernova explosion in which pulsar B was formed [82]. This is in agreement with the small eccentricity observed, which suggests little mass ejection during the second collapse [83]. The study of the GR effects is simplified by the slow proper motion of the binary system, because this motion contributes to the observed increase in $\dot{P}_{\rm b}$ due to the relative motion, leading to a negative angular acceleration of the system [84, 85]. This contribution to the orbital period increase is much smaller than 1%.

The most accurately measured parameters in Table 2, including the mass ratio $R$, the periastron advance rate $\dot{\omega}$, and the Shapiro delay parameter $s$, can already be used to test the theory of gravity (Fig. 5). Assuming GR to be the correct gravity theory, we use Eqn (14) to obtain the total mass of the system, which in combination with the known mass ratio enables us to determine $M_{\rm A} = 1.338 \pm 0.001 M_\odot$ and $M_{\rm B} = 1.249 \pm 0.001 M_\odot$. Using these mass values, we can use GR to calculate the Shapiro delay parameter $s$ and compare it with the observed value. It was found in Ref. [71] that $s^{\rm GR}/s^{\rm obs} = 1.0002^{+0.0011}_{-0.0006}$. Thus, GR passes this test at a level of 0.1%, which is by now the best test of the validity of GR in the strong-field limit.

### 9.3 Further prospects

In a few years, the extended observational time and instrumental progress will ensure measurements of additional PK parameters [69], including higher-order PK terms. In particular, it will be possible to study the effect of neutron star spins on their orbital motion, which is impossible in Newtonian theory. This effect should be observed as secular [77] and periodic [86] changes in the observed value of $\dot{\omega}$. For the double pulsar J0737–3039, this effect should be an order of magnitude stronger than for PSR 1913+16, being about $2 \times 10^{-4}$ deg yr$^{-1}$ for pulsar A. Measuring this effect gives for the first time the possibility of determining the neutron star moment of inertia [79, 87].

Sufficiently accurate measurements of the Shapiro delay parameter $s$ and the mass ratio $R$ allow the value of $\dot{\omega}_{\rm exp}$ to be found as the intersection point. This value can be compared with the observed one $\dot{\omega}_{\rm obs}$, which can be represented as [87]

$$\dot{\omega}_{\rm obs} = \dot{\omega}_{\rm 1PN} \left[ 1 + \Delta\dot{\omega}_{\rm 2PN} - g^{\rm A} \Delta\dot{\omega}^{\rm A}_{\rm SO} - g^{\rm B} \Delta\dot{\omega}^{\rm B}_{\rm SO} \right], \qquad (21)$$

were the last two terms are due to spin contributions from two pulsars, with $g^{\rm A,B}$ depending on the geometry and $\Delta\dot{\omega}^{\rm A,B}_{\rm SO}$ being due to the relativistic spin–orbit interaction, formally at the 1PN level. However, in the double-pulsar system, these terms turn out to be comparable to the 2PN terms [86], and therefore taking them into account is needed only when higher-order terms in the $\dot\omega$ expansion are considered. The value $\Delta\dot\omega_{SO} \propto I/PM^2$ [87], and hence the precise masses $M$ enable the measurement of the neutron star moment of inertia $I$. This yields a unique possibility of studying the equation of state of matter with superhigh density and pressure.

### 9.4 The past and future of pulsars in a binary system

The emission of gravitational waves results in the ultimate coalescence of the components after about 85 mln years. The discovery of this system significantly increases the estimated probability of detecting binary neutron star coalescences by their gravitational wave emission using gravitational wave detectors [68]. The simultaneous timing of both pulsars in their motion in the common gravity field increases the accuracy of GR tests and measurements of gravitational wave emission. The properties of pulsars in this binary system perfectly fit the evolutionary formation scheme for binary pulsars suggested in Ref. [6]. According to this scenario, the 23 ms pulsar was recycled by accretion and its magnetic field was simultaneously screened by the accreted plasma. The second pulsar with the period 2.8 s resulted later from the second supernova explosion that did not destroy the binary system. Therefore, the millisecond pulsar in this binary should be older and have a weaker magnetic field. Indeed, the magnetic field of the 23 ms pulsar was estimated to be $B_{ms} = 6.3 \times 10^9$ G and its characteristic age is $\tau = 210$ mln years. The ordinary 2.8 s pulsar has the field $B_n = 1.6 \times 10^{12}$ G and its age is $\tau = 50$ mln years [2]. The masses of neutron stars in this system are $1.34 M_\odot$ and $1.25 M_\odot$ for the 23 ms and 2.8 s pulsars, respectively.

Knowing the current parameters of this system allows tracing its past and future evolution [83], because gravitational radiation is the only physical process changing the binary system parameters. Magnetic stellar wind and tidal interaction acting in ordinary stars are not important in the binary neutron star system. Changes to the binary major semiaxis $a$ and eccentricity $e$ in a system with masses $m_1$ and $m_2$, averaged over the orbital period, are given by the formulas [88]

$$\frac{da}{dt} = -\frac{64}{5}\frac{m_1 m_2 (m_1+m_2)}{a^3(1-e^2)^{7/2}}\left(1+\frac{73}{24}e^2+\frac{37}{96}e^4\right), \quad (22)$$

$$\frac{de}{dt} = -\frac{304}{15}\frac{m_1 m_2 (m_1+m_2)e}{a^4(1-e^2)^{5/2}}\left(1+\frac{121}{304}e^2\right). \quad (23)$$

The current values of the parameters of the double pulsar system are [2]

$m_1 = 1.34 M_\odot$ (millisecond pulsar),
$m_2 = 1.25 M_\odot$, $e_0 = 0.0878$, (24)
$P_{b0} = 0.102$ d $= 8.83 \times 10^3$ s, $a_0 = 8.8 \times 10^{10}$ cm.

The last quantity in Eqn (24) is derived from Kepler's third law,

$$\frac{a^3}{P_b^2} = \frac{G(m_1+m_2)}{4\pi^2}, \quad (25)$$

for known stellar masses and binary period $P_{b0}$. Integration of system (22), (23) yields the functions $a(t)$, $e(t)$, and $P_b(t)$

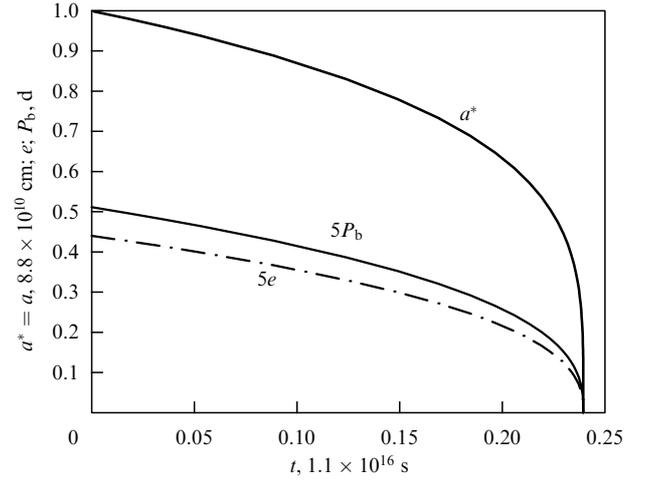

**Figure 6.** Evolution of the binary system parameters (the large semi-major axis $a$, orbital period $P_b$ and eccentricity $e$) due to gravitational wave emission [83]. d

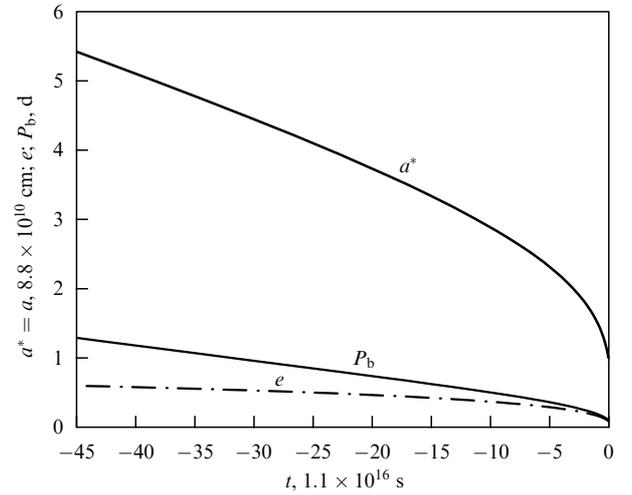

**Figure 7.** Possible past evolution of the binary system parameters (the large semi-major axis $a$, orbital period $P_b$ and eccentricity $e$) due to gravitational wave emission [83].

(Fig. 6 for the future, Fig. 7 for the past). The lifetime of the system before the coalescence (Fig. 6) is 84 mln years in accordance with the estimate in Ref. [2]. Integration into the past shows that over the characteristic lifetime of the second (50 mln years) and millisecond (210 mln years) pulsars, the binary parameters change insignificantly (Fig. 7). At the same time, it is very difficult to imagine why the orbital eccentricity $e$ did not increase after the collapse and supernova explosion of the second companion, as is observed in other binary neutron star systems, which have $e > 0.18$, for example, $e = 0.617$ (PSR 1913+16) and $e = 0.68$ (PSR 2121+11C) [28, 51]. Therefore, one should accept that either the pulsar ages, as suggested by their rotation deceleration measurements, are strongly underestimated, or the second collapse in this binary system indeed occurred virtually without mass ejection and did not increase the eccentricity. It is likely that the complete alignment of orbital and spin periods occurred in such a close binary system at the pre-supernova stage. In that case, the spin period of the second neutron star immediately after formation would be 10–100 ms and its energy would be insufficient for the magnetorotational explosion [89] and the mass ejection would be insignificant. The tidal relaxation that was intensive in this system in the past is also suggested by both pulsars being observable, which is possible for similar beaming properties. If the system, which evolves only due to gravitational wave emission, were old, then its parameters 7 bln years ago would be $a = 3.76\, a_0$, $P = 0.73$ days, and $e = 0.5$.

## 10. Binary neutron star coalescences: gravitational wave and gamma-ray bursts

The main burst of gravitational waves during the coalescence of binary neutron stars (BNS) is generated immediately before their merging. BNSs are the principal potential sources of gravitational waves. At present, the most sensitive detectors include the ground-based laser interferometers LIGO [90] (Laser Interferometer Gravitational-Wave Observatory), GEO [91] and VIRGO [92]. Estimates of the merging rates of relativistic binary objects (neutron stars and black holes) are very important in choosing the strategy of construction of gravitational wave detectors [93, 94]. These estimates are related to the merging rates in the local volume from which the gravitational wave signal can be detected and depend on the source power and detector sensitivity. These estimates can be done by using a theoretical approach based on stellar evolution calculations and the observed frequency of supernova explosions. The empirical approach, which is based on the properties of a known BNS and accounts for selection effects, appears to be more reliable. However, estimates in the latter case are also uncertain by more than two orders of magnitude [95] in view of the small number of known close BNSs. The lifetime of the BNS PSR J0737−3039 before merging (about 85 mln years) is about 3.5 times shorter than that of PSR 1913+16, which increases the expected coalescence rate [68]. In Ref. [95], in addition to considering the properties of the double-pulsar system, a modified statistical method [96] was applied to estimate the expected coalescence rate. This method enabled the authors to construct the curve of probability of the expected coalescence rate based on the properties of four known close BNSs with the coalescence times shorter than the cosmological one. In addition to the BNSs discussed above, these systems include RPs B1534+12 and B2127+11C (Table 3).

BNS B2127+11C enters the globular cluster M15, and as was shown in Ref. [97], the merging rate of such a BNS in the galaxy is negligibly small in comparison with the BNS field, and therefore the analysis of the BNS coalescence rate in Ref. [95] was in fact performed by considering the properties of three other BNSs. As shown in Ref. [96], the form of the BNS coalescence rate probability is weakly dependent on the parameters. At the same time, the rate at the maximum of the probability curve is strongly determined by the radio pulsar luminosity function slope, as well as by the physical minimum of the radio pulsar luminosity. The constraints on these parameters were obtained in Ref. [98] and used in calculations in Ref. [95]. The total lifetime of pulsar J0737−3039, defined as the sum $\tau_c + \tau_g$, was taken to be $100 + 85 = 185$ mln years [68].

### 10.1 Galactic binary neutron star coalescence rate and gravitational-wave pulse registration

For the model of pulsar evolution in Ref. [98], the mean galactic merging rate of BNS systems is obtained in Ref. [95] to be $\mathcal{R} = 83$ Myr$^{-1}$. The 68%- and 95%-confidence level intervals are 40–140 and 20–290 Myr$^{-1}$, respectively. The expected detection rate of a gravitational-wave pulse from neighboring galaxies is 0.035 and 190 events per year for the initial (the detection limit 20 Mpc) and advanced (the detection limit 350 Mpc) LIGO interferometers, respectively. The corresponding 95%-confidence intervals are 0.007–0.12 and 40–660 events per year, respectively [95].

The discovery of the double pulsar J0737-3039 increased $\mathcal{R}$ by 6.4 times compared to earlier calculations [96] because it dominates in computing the total probability, as seen in Fig. 8.

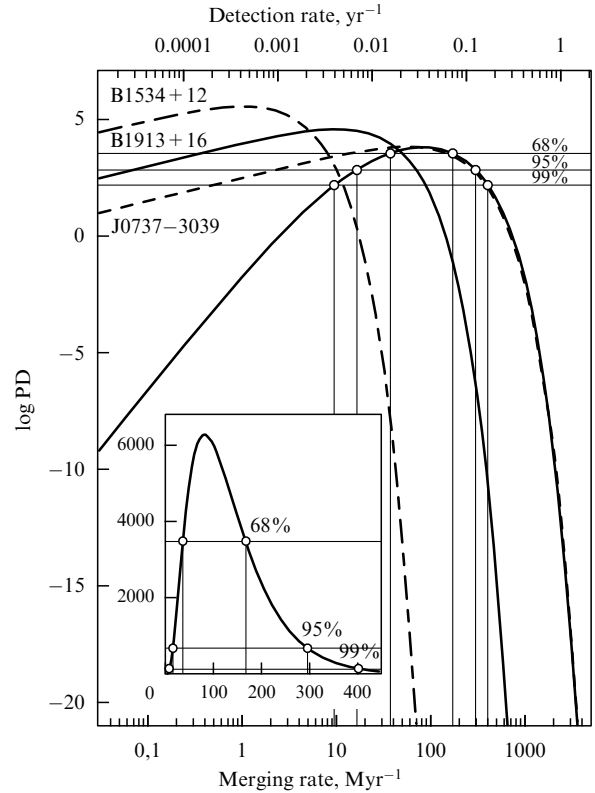

**Figure 8.** The logarithm of the probability density (log PD) characterizing the BNS rate in the galaxy (the bottom axis) and the predicted detection rate by the initial LIGO interferometer (the upper axis). The calculations use the data on three BNSs with restrictions from Ref. [98] taken into account. The solid curve shows the total detection probability and the dashed lines correspond to individual pulsars. In the inset, the total probability density and the 68, 95, and 99% confidence levels are shown on the linear scale [95].

**Table 3.** Parameters of three close BNSs: the pulsar period $P$, the orbital period $P_b$, the orbital eccentricity $e$, the characteristic pulsar age $\tau_c = P/2\dot{P}$, and the expected coalescence time $\tau_g$ due to gravitational wave emission [28, 95].

| Parameter | Pulsar | | |
|---|---|---|---|
| | B1534+12 | B1913+16 | B2127+11C |
| $P$, ms | 37.9 | 59.0 | 30.5 |
| $P_b$, d | 0.42 | 0.32 | 0.34 |
| $e$ | 0.27 | 0.62 | 0.68 |
| $\tau_c$, $10^8$ yrs | 2.5 | 1.1 | 0.97 |
| $\tau_g$, $10^8$ yrs | 27 | 3.0 | 2.2 |

Such an increase is firstly due to a larger expected galactic number of a BNS-like J0737–3039 (1700) compared to a BNS-like B1913+16 (700) or B1534+12 (500), and, secondly, due to the shorter total life time of J0737–3039 (185 mln years) compared to the other two (365 mln years and 2.9 bln years for B1913+16 and B1534+12, respectively). Examination of a broader class of evolutionary models of pulsars showed that in all cases, accounting for the double pulsar J0737–3039 increases the BNS merging rate by 6–7 times, although the rates can differ by more than 50 times in individual cases [95].

## 10.2 Binary neutron star coalescences and gamma-ray bursts

Coalescences of binary neutron stars are accompanied by a huge energy release, and therefore they have long been regarded as possible sources of gamma-ray bursts [93, 99]. The cosmological nature of gamma-ray bursts (GRBs), as established by measuring redshifts of lines in the spectra of their afterglows, points to an enormous energy release that hardly can result from such coalescences. Neutron star mergings were numerically modeled in Refs [100, 101]. In these models, gamma-ray emission is produced by $(v, \bar{v})$ annihilation, and the energy released is found to be insufficient to explain the most powerful cosmological GRBs, even under the assumption of strong collimation of gamma-ray emission. The energy emitted only in the optical afterglow of GRB 990123 [102, 103] is an order of magnitude higher than the total electromagnetic energy in this model.

The cosmological origin has so far been established only for a certain class of fairly long GRBs (several seconds) with complex multi-peak light curves. About 40% of GRBs exhibit one short pulse of gamma-ray emission lasting less than 1 s. It is quite possible that short GRBs are less energetic and can be due to BNS coalescence.

When the masses of neutron star components are significantly different, their coalescing can be accompanied by an intensive mass transfer to the more massive star. When the mass of the 'light' neutron star decreases to about $0.09 M_\odot$, the hydrostatic equilibrium is violated and the star starts losing mass in outflow until complete disruption. This happens because neutron stars of very low mass, due to peculiarities of the equation of state, have positive total energy. Because all characteristic times in neutron stars are shorter than a fraction of a second, the outflow and the subsequent disruption can be viewed by an external observer as different stages of one explosion. This kind of evolution of a BNS was first investigated in Ref. [104]. The authors also noted that the disruption of a low-mass neutron star can probably be accompanied by a gamma-ray burst due to a nuclear reaction, considered earlier in Ref. [105]. In the application to the GRB problem, this reaction was considered to occur in matter consisting of superheavy neutron-overloaded nuclei that come to the neutron star surface from deeper layers due to a star quake.

## 11. Conclusion

While it took 7 years after the discovery of radio pulsars for the first pulsar in a close binary system to be found, it took 30 years for the first double-pulsar system to be discovered. This indicates that such binaries occur fairly rarely and apparently is due to a high probability that most neutron stars in binaries have different beams that are simultaneously observed from the earth. Such decoupling could be a result of an asymmetric supernova explosion, during which the recoil effects could change the spin vectors of neutron stars and break their coupling that could have been established by tidal interaction at earlier quiet evolutionary stages. The asymmetry may appear during the magnetorotational supernova explosion with a collapsing core, when the spin axis is not coincident with the magnetic field axis and where, as shown in Ref. [106], the development of the magneto-rotational instability is essential.

The strikingly slow proper motion of the double-pulsar system does not seem to be accidental. This could be due to the small anisotropy of the explosion or a small recoil effect and weak decoupling of angular momenta of neutron stars, which enables both pulsars to be simultaneously observed from the earth. This may suggest that such objects should be sought near the galactic plane, where pulsars with slow proper motion and weak supernova explosion asymmetry are left.

Measurements of five PK parameters in the double-pulsar system enabled highly precise determination of the mass ratio, and further measurements of pulse arrival times from both pulsars will strongly increase the measurement accuracy of these parameters, while permitting observational investigation of other relativistic effects. This will significantly increase the reliability of GR, or provide evidence for its modifications.

The author would like to thank B V Komberg for discussions and useful notes. The work was supported by the RFBR grant 05-02-17697A.

## References

1. Hulse R A, Taylor J H *Astrophys. J. Lett.* **195** L51 (1975)
2. Lyne A G et al. *Science* **303** 1153 (2004)
3. Hewish A et al. *Nature* **217** 709 (1968)
4. Shwartsman V F *Astron. Zh.* **48** 438 (1971) [*Sov. Astron.* **15** 342 (1971)]
5. Johnston S et al. *Astrophys. J. Lett.* **387** L37 (1992)
6. Bisnovatyi-Kogan G S, Komberg B V *Astron. Zh.* **51** 373 (1974) [*Sov. Astron.* **18** 217 (1974)]
7. Bisnovatyi-Kogan G S *Priroda* (2) 15 (1995)
8. Herring C, Pines D *Phys. Today* **28** (11) 46 (1975)
9. Taylor J H et al. *Astrophys. J. Lett.* **206** L53 (1976)
10. Bisnovatyi-Kogan G S, Komberg B V, Preprint Pr-269 (Moscow: Space Research Institute of the Academy of Sciences of the USSR, 1976)
11. Bisnovatyi-Kogan G S, Komberg B V *Pis'ma Astron. Zh.* **2** 338 (1976) [*Sov. Astron. Lett.* **2** 130 (1976)]
12. Bisnovatyi-Kogan G S *Priroda* (3) 100 (1975)
13. van den Heuvel E P J *Science* **303** 1143 (2004)
14. Ginzburg V L, Zheleznyakov V V, Zaitsev V V *Astrophys. Space Sci.* **4** 464 (1969)
15. Landau L D, Lifshitz E M *Teoriya Polya* (The Classical Theory of Fields) (Moscow: Nauka, 1962) [Translated into English (Oxford: Pergamon Press, 1975)]
16. Pacini F *Nature* **216** 567 (1967)
17. Malov I F *Radiopul'sary* (Radiopulsars) (Moscow: Nauka, 2004)
18. Goldreich P, Julian W H *Astrophys. J.* **157** 869 (1969)
19. Lyne A G, Graham-Smith F *Pulsar Astronomy* 2nd ed. (Cambridge: Cambridge Univ. Press, 1998)
20. Trimble V, Rees M, in *The Crab Nebula: Proc. of the IAU Symp. No. 46, Jodrell Bank, England, August 5–7, 1970* (Eds R D Davies, F Graham-Smith) (Dordrecht: Reidel, 1971) p. 273
21. Wijnands R, van der Klis M *Nature* **394** 344 (1998)
22. Ruderman M, Shaham J *Comments Mod. Phys. C: Comments Astrophys.* **10** (9) 15 (1983)
23. Galloway D K et al. *Astrophys. J.* **622** L45 (2005); astro-ph/0501064
24. Wijnands R, astro-ph/0501264
25. Bisnovatyi-Kogan G S *Astrofizika* **31** 567 (1989); **32** 193 (1990)

26. Bisnovatyi-Kogan G S *Astrofizika* **32** 313 (1990) [*Astrophys.* **32** 176 (1990)]
27. Lovelace R V E, Romanova M M, Bisnovatyi-Kogan G S *Mon. Not. R. Astron. Soc.* **275** 244 (1995)
28. Lorimer D R *Living Rev. Relativity* **4** (6) 5 (2001); astro-ph/0104388
29. Trümper J et al. *Astrophys. J. Lett.* **219** L105 (1978)
30. Baushev A N, Bisnovatyi-Kogan G S *Astron. Zh.* **76** 283 (1999) [*Astron. Rep.* **43** 241 (1999)]
31. Sheffer E K et al. *Astron. Zh.* **69** 82 (1992) [*Sov. Astron* **36** 41 (1992)]
32. Scott D M, Leahy D A, Wilson R B *Astrophys. J.* **539** 392 (2000)
33. Ramsay G et al. *Mon. Not. R. Astron. Soc.* **337** 1185 (2002)
34. van den Heuvel E P J, Bitzaraki O *Astron. Astrophys.* **297** L41 (1995)
35. Bisnovatyi-Kogan G S *Riv. Nuovo Cimento, Ser. 3* **2** (1) 1 (1979)
36. Choudhuri A R, Konar S *Mon. Not. R. Astron. Soc.* **332** 933 (2002)
37. Cheng K S, Zhang C M *Astron. Astrophys.* **337** 441 (1998)
38. Cheng K S, Zhang C M *Astron. Astrophys.* **361** 1001 (2000)
39. Choudhuri A R, Konar S *Current Sci.* **86** 444 (2004)
40. Bisnovatyi-Kogan G S, Fridman A M *Astron. Zh.* **46** 721 (1969) [*Sov. Astron.* **13** 566 (1970)]
41. Amnuel P R, Guseinov O H *Astron. Tsirk.* (524) (1969)
42. Shapiro S L, Teukolsky S A *Black Holes, White Dwarfs, and Neutron Stars: the Physics of Compact Objects* (New York: Wiley, 1983)
43. Konar S, Choudhuri A R *Mon. Not. R. Astron. Soc.* **348** 661 (2004)
44. Lovelace R V E, Romanova M M, Bisnovatyi-Kogan G S *Astrophys. J.* **625** 957 (2005)
45. Michel F C *Mon. Not. R. Astron. Soc.* **267** L4 (1994)
46. Muslimov A, Page D *Astrophys. J. Lett.* **440** L77 (1995)
47. Tutukov A V, Yungelson L R *Nauchnye Informatsii* **27** 86 (1973)
48. Bhattacharya D, van den Heuvel E P J *Phys. Rep.* **203** 1 (1991)
49. Bisnovatyi-Kogan G S, Moiseenko S G *Astron. Zh.* **69** 563 (1992) [*Sov. Astron.* **36** 285 (1992)]
50. Bisnovatyi-Kogan G S *Astron. Astrophys. Trans.* **3** 287 (1993)
51. Faulkner A J et al. *Astrophys. J.* **618** L119 (2004); astro-ph/0411796
52. Camilo F, Rasio F A, in *Binary Radio Pulsars* (Astron. Soc. of the Pacific Conf. Ser., Vol. 328, Eds F A Rasio, I H Stairs) (San Francisco: ASP, 2005) p. 147
53. Kluzniak W et al. *Nature* **334** 225 (1988)
54. Shklovskii I S *Comments Mod. Phys. C: Comments Astrophys.* **9** (6) 261 (1982)
55. Bisnovatyi-Kogan G S, Romanova M M *Astron. Zh.* **60** 900 (1983) [*Sov. Astron.* **27** 519 (1983)]
56. Rasio F A, astro-ph/0212211
57. Rappaport S, Putney A, Verbunt F *Astrophys. J.* **345** 210 (1989)
58. Gurevich L E, Levin B Yu *Astron. Zh.* **27** 273 (1950)
59. Aarseth S J, Lecar M *Annu. Rev. Astron. Astrophys.* **13** 1 (1975)
60. Zapolsky H S, Salpeter E E *Astrophys. J.* **158** 809 (1969)
61. Tutukov A V *Nauchnye Informatsii* **11** 69 (1969)
62. Spitzer L (Jr.) *Dynamical Evolution of Globular Clusters* (Princeton, NJ: Princeton Univ. Press, 1987) [Translated into Russian (Moscow: Mir, 1990)]
63. Damour T, Deruelle N *Ann. Inst. Henri Poincaré Sect. A* **44** 263 (1986)
64. Brumberg V A et al. *Pis'ma Astron. Zh.* **1** 5 (1975) [*Sov. Astron. Lett.* **1** 2 (1975)]
65. Martynov D Ya *Kurs Obshchei Astrofiziki* (General Astrophysics) 4th ed. (Moscow: Nauka, 1988)
66. Weisberg J M, Taylor J H, in *Radio Pulsars* (Astron. Soc. of the Pacific Conf. Ser., Vol. 302, Eds M Bailes, D J Nice, S E Thorsett) (San Francisco, Calif.: Astron. Soc. of the Pacific, 2003) p. 93
67. Taylor J H (Jr) *Rev. Mod. Phys.* **66** 711 (1994)
68. Burgay M et al. *Nature* **426** 531 (2003)
69. Damour T, Taylor J H *Phys. Rev. D* **45** 1840 (1992)
70. Stairs I H et al. *Astrophys. J.* **581** 501 (2002)
71. Kramer M et al., astro-ph/0503386
72. McLaughlin M A et al. *Astrophys. J. Lett.* **613** L57 (2004)
73. Kaspi V M et al. *Astrophys. J. Lett.* **613** L137 (2004)
74. McLaughlin M A et al. *Astrophys. J. Lett.* **616** L131 (2004)
75. Taylor J H *Philos. Trans. R. Soc. London Ser. A* **341** 117 (1992)
76. Damour T, Ruffini R *C.R. Acad. Sci. Ser. A Sci. Math.* (Paris) **279** 971 (1974)
77. Barker B M, O'Connell R F *Astrophys. J.* **199** L25 (1975)
78. Stairs I H, Thorsett S E, Arzoumanian Z *Phys. Rev. Lett.* **93** 141101 (2004)
79. Kramer M *Astrophys. J.* **509** 856 (1998)
80. Jenet F A, Ransom S M *Nature* **428** 919 (2004)
81. Cordes J M, in *Milky Way Surveys: The Structure and Evolution of our Galaxy* (Astron. Soc. of the Pacific Conf. Ser., Vol. 317, Eds D Clemens, R Y Shah, T Brainerd) (San Francisco, Calif.: Astron. Soc. of the Pacific, 2004) p. 211
82. Piran T, Shaviv N J *Phys. Rev. Lett.* **94** 051102 (2005)
83. Bisnovatyi-Kogan G S, Tutukov A V *Astron. Zh.* **81** 797 (2004) [*Astron. Rep.* **48** 724 (2004)]
84. Shklovskii I S *Astron. Zh.* **46** 715 (1969) [*Sov. Astron.* **13** 562 (1969)]
85. Damour T, Taylor J H *Astrophys. J.* **366** 501 (1991)
86. Wex N *Class. Quantum Grav.* **12** 983 (1995)
87. Damour T, Schäfer G *Nuovo Cimento* **101** 127 (1988)
88. Lightman A P et al. *Problem Book in Relativity and Gravitation* (Princeton, NJ: Princeton Univ. Press, 1975) [Translated into Russian (Moscow: Mir, 1979)]
89. Bisnovatyi-Kogan G S *Stellar Physics* Vol. 2 *Stellar Evolution and Stability* (Berlin: Springer, 2002)
90. Abramovici A et al. *Science* **256** 325 (1992)
91. Danzmann K et al., in *First Edoardo Amaldi Conf. on Gravitational Wave Experiments, Frascati, Rome, Italy, 14 – 17 June 1994* (Edoardo Amaldi Foundation Ser., Vol. 1, Eds E Coccia, G Pizzella, F Ronga) (Singapore: World Scientific, 1995) p. 100
92. Caron B et al. *Nucl. Phys. B: Proc. Suppl.* **54** 167 (1997)
93. Lipunov V M, Postnov K A, Prokhorov M E *Astrophys. Space Sci.* **252** 401 (1997)
94. Cutler C, Thorne K, in *Proc. of the 16th Conf. on General Relativity & Gravitation, Durban, South Africa, 15 – 21 July 2001* (Eds N T Bishop, S D Maharaj) (Singapore: World Scientific, 2002) p. 72
95. Kalogera V et al. *Astrophys. J. Lett.* **601** L179; **614** L137 (2004)
96. Kim C, Kalogera V, Lorimer D R *Astrophys. J.* **584** 985 (2003)
97. Phinney E S *Astrophys. J. Lett.* **380** L17 (1991)
98. Cordes J M, Chernoff D F *Astrophys. J.* **482** 971 (1997)
99. Lipunov V M et al. *Astrophys. J.* **454** 593 (1995)
100. Ruffert M, Janka H-Th *Astron. Astrophys.* **338** 535 (1998)
101. Ruffert M, Janka H-Th *Astron. Astrophys.* **344** 573 (1999)
102. Akerlof C et al. *Nature* **398** 400 (1999)
103. Kulkarni S R et al. *Nature* **398** 389 (1999)
104. Blinnikov S I et al. *Pis'ma Astron. Zh.* **10** 422 (1984) [*Sov. Astron. Lett.* **10** 177 (1984)]
105. Bisnovatyĭ-Kogan G S, Chechetkin V M *Usp. Fiz. Nauk* **127** 263 (1979) [*Sov. Phys. Usp.* **22** 89 (1979)]
106. Ardeljan N V, Bisnovatyi-Kogan G S, Moiseenko S G *Mon. Not. R. Astron. Soc.* **359** 333 (2005)